\documentclass{PoS}
\usepackage{ragged2e}
\usepackage{colordvi}
\usepackage{multirow}
\usepackage{amsmath,amsfonts,epsfig,cite,graphics}
\usepackage{amssymb,bbm}
\usepackage{mathrsfs}
\usepackage{amsbsy}
\usepackage{latexsym}
\usepackage{graphicx}
\usepackage{psfrag}
\usepackage{slashed}
\usepackage{macros_static}
\newcommand{\qvec}{\vec{q}}

\newcommand{\<}{\langle}
\renewcommand{\>}{\rangle}
\renewcommand{\l}{\left}
\renewcommand{\r}{\right}

\newcommand{\idnty}{\hbox{1$\!\!\!$1}}

\def\eq#1{Eq.~(\ref{#1})}
\def\fig#1{Fig. \ref{#1}}

\def\sec#1{Section \ref{#1}}

\def\cyp{a}
\def\cyi{b}
\def\was{c}
\def\nic{d}
\def\hisk{e}

\title{\vspace{-25mm} \hfill{\small DESY 15-203}\\[5mm]
The electric dipole moment of the neutron from $N_f=2+1+1$ twisted mass fermions.\vspace{-5mm}}

\ShortTitle{The nEDM with $N_f=2+1+1$ twisted mass fermions.}

\author{ C.~Alexandrou$^{(\cyp, \cyi)}$, \speaker{A.~Athenodorou}$^{ \ (\cyp, \cyi)}$, M.~Constantinou$^{(\cyp, \cyi)}$, K.~Hadjiyiannakou$^{(\cyp, \cyi, \was)}$, K.~Jansen$^{(\nic)}$, G.~Koutsou$^{(\cyi)}$, K.~Ottnad$^{(\cyp, \hisk)}$, M.~Petschlies$^{(\cyi, \hisk)}$ \\ \\
\llap{$^{(\cyp)}$} Department of Physics, University of Cyprus, P.O. Box 20537,
 	1678 Nicosia, Cyprus\\
\llap{$^{(\cyi)}$} Computation-based Science and Technology Research Center, The Cyprus Institute, 20 Kavafi Str., Nicosia 2121, Cyprus\\
\llap{$^{(\was)}$} Department of Physics, The George Washington University, Washington, DC 20052, USA\\
 \llap{$^{(\nic)}$} NIC, DESY, Platanenallee 6, D-15738 Zeuthen, Germany\\
\llap{$^{(\hisk)}$} Helmholtz-Institut f\"ur Strahlen- und Kernphysik (Theorie) and Bethe Center for Theoretical Physics, Universit\"at Bonn, 53115 Bonn, Germany \vspace{5mm} \\
E-mail:\email{alexand@ucy.ac.cy}, \email{athenodorou.andreas@ucy.ac.cy}, \email{marthac@ucy.ac.cy}, \email{hadjigiannakou.kyriakos@ucy.ac.cy}, \email{Karl.Jansen@desy.de}, \email{g.koutsou@cyi.ac.cy}, \email{ottnad@hiskp.uni-bonn.de}, \email{m.petschlies@cyi.ac.cy}}

\abstract{We extract the neutron electric dipole moment (nEDM) $ \vert \vec{d}_n \vert$ on configurations produced with $N_f=2+1+1$ twisted mass fermions with lattice spacing of $a \simeq 0.082$fm and a light quark mass that corresponds to $M_{\pi} \simeq 373$ MeV. We do so by evaluating the $CP$-odd form factor $F_3$ for small values of the $CP$-violation parameter $\theta$ in the limit of zero momentum transfer. This limit is extracted using the usual parametrization but in addition  position space methods. The topological charge is computed via cooling and gradient flow using the Wilson, Symanzik tree-level improved and Iwasaki actions for smoothing. We obtain consistent results for all choices of smoothing procedures and methods to extract $F_3$ at zero momentum transfer. For the ensemble analyzed we find a value of nEDM of $\vert \vec{d}_n \vert/ \theta = −0.045(6)(1) {\rm e} \cdot {\rm fm}$.}

\PACS{12.38.-t, 12.38.Gc, 12.60.-i, 13.40.Gp \hspace{\stretch{1}} }

\FullConference{The ${33^{\rm rd}}$ International Symposium on Lattice Field Theory\\
         July 14-18 2015\\
         Kobe International Conference Center, Kobe, Japan}

\begin{document}
\section{Introduction}
\label{sec:introduction}
A non-vanishing neutron Electric Dipole Moment (nEDM) would signal $CP$-violation. Up to date, no finite nEDM has been reported in experiments and the best experimental bound~\cite{Baker:2006ts}, $\vert \vec{d}_N \vert  < 2.9 \times 10^{-13} e \cdot {\rm fm}$, is several orders of magnitude below what one expects from $CP$-violation induced by weak interactions~\cite{Pospelov:2005pr}.
This makes nEDM searches an interesting probe for Beyond the Standard Model (BSM) physics.

In strong dynamics, a non-vanishing  nEDM can be induced by a Lagrangian density consisting of the ordinary QCD part, plus, a $CP$-violating interaction (Chern-Simons) term
\begin{eqnarray}
\vspace{-0.25cm}
{\cal L}_{\rm QCD} \left( x \right)= \frac{1}{2 g^2} {\rm Tr} \left[ G_{\mu \nu} \left( x \right) G_{\mu \nu} \left( x \right) \right] + 
\sum_{f} {\overline \psi}_{f} \left( x \right) (\gamma_{\mu} D_{\mu} + m_f) \psi_{f}\left( x \right)- i \theta q \left( x \right)\, ,
\label{eq:CP_odd_action}
\vspace{-0.25cm}
\end{eqnarray}
(in Euclidean space) where the $\theta$-parameter controls the strength of the $CP$-breaking and $q(x)$ the topological charge density defined as $q \left( x \right) = \frac{1}{ 32 \pi^2}  \epsilon_{\mu \nu \rho \sigma} {\rm Tr} \left[ G_{\mu \nu} \left( x \right) G_{\rho \sigma} \left( x \right) \right]$. Results from several model studies~\cite{Pospelov:2005pr} suggest that $\theta$ is small $\left(\theta \lesssim {\cal O} \left( 10^{-10} - 10^{-11} \right)\right)$.

The magnitude of the nEDM at leading order in $\theta$ is expressed as $\vert \vec{d}_N \vert =  \theta \lim_{Q^2 \to 0} {\vert F_3(Q^2) \vert}/{2 m_N}\,$ \cite{Pospelov:2005pr}
where $m_N$ denotes the mass of the neutron, $Q^2$ the four-momentum 
transfer in Euclidean space and $F_3(Q^2)$ the $CP$-odd form factor. 
Hence, we can calculate the nEDM by evaluating the zero momentum transfer limit of $F_3(Q^2)$. The $CP$-odd nucleon matrix element gives access to $F_3(Q^2)$. 
\section{The $CP$-odd nucleon matrix element}
\label{sec:cp_odd_matrix_element}
The expectation value of an operator ${\cal O}$ in a theory with broken $CP$-symmetry and, thus, in $\theta \ne 0$ vacuum, can be obtained by using the path integral with Lagrangian density
given in \eq{eq:CP_odd_action}. Such expectation value for small values of $\theta$ can be expressed and expanded as
\begin{eqnarray}
    \< {\cal O}(x) \>_{\theta}= \frac{1}{Z_\theta} \hspace{-0.05cm} \int \hspace{-0.05cm} d[U] d[\psi_f] 
    d[\bar{\psi}_f] \hspace{-0.05cm} ~ \hspace{-0.05cm} {\cal O}(x) ~ e^{-S_{\rm QCD}+i{\theta} {\cal Q}}  \hspace{-0.05cm} = \hspace{-0.05cm} \< {\cal O}(x) \>_{\theta=0}+ i{\theta}\,\left\< {\cal O}(x) {\cal Q} \right\>_{\theta=0}+{O}(\theta^2)\,,
    \label{eq:vev}
\end{eqnarray}
where  ${\cal Q}=\int  d^4x \,q(x)$ is the topological charge. To extract $F_3$ at $Q^2=-(p_f-p_i)^2$ we consider the matrix element of the electromagnetic current $J_\mu^{\rm em}$ in the $\theta$-vacuum $\langle N ({\vec p}_f,s_f)\vert J_\mu^{\rm em}\vert N ({\vec p}_i,s_i)\rangle_{\theta} = \bar u_N ({\vec p}_f,s_f)\, W_\mu^\theta (Q)\, u_N ({\vec p}_i,s_i)$ where $p_f$ ($p_i$) and $s_f$ ($s_i$) are the momentum and spin of the
final (initial) nucleon state $N$ and
\bea
W_\mu^\theta (Q) = F_1(Q^2) - i\,\frac{F_2(Q^2)}{2m_N}\,Q_\nu\,\sigma_{\nu\mu} + i \theta \left(  -i\, \frac{F_3(Q^2)}{2m_N}\, Q_\nu\,\sigma_{\nu\mu}\gamma_5
+ F_A(Q^2)\left(Q_\mu \slashed{Q} - \gamma_\mu\,Q^2\right)\gamma_5\,  \right),
\label{eq:forms}
\eea
at leading order in $\theta$. $F_1(Q^2)$ and $F_2(Q^2)$ are the $CP$-even Pauli and
Dirac form factors and $F_3(Q^2)$ and $F_A(Q^2)$ the $CP$-odd electric dipole and anapole form factors respectively. 
To extract the nEDM we expand the 3-point(pt) function according to \eq{eq:vev} as $G_{\rm 3pt}^{\mu, \left(\theta \right)} (\qvec, t_f, t_i, t )=G_{\rm 3pt}^{\mu,(0)} (\qvec, t_f, t_i, t ) + i \,\theta \, G_{\rm 3pt, {\cal Q}}^{\mu,(0)} (\qvec, t_f, t_i, t ) +{\it O} \left( \theta^2 \right)$ where $G_{\rm 3pt, {\cal Q}}^{\mu, \left(0\right)} (\qvec, t_f, t_i, t ) = $ $\langle J_N ({\vec p}_f,t_f) J^{\rm em}_\mu(\vec q, t)  {\overline J}_N({\vec p}_i,t_i) {\cal Q} \rangle$ is the vacuum expectation value for $\theta=0$ and $\vec{q}\equiv\vec{Q}$. By spin-projecting the above 3-pt function with $\Gamma_k=\frac{i}{4}(\idnty+\gamma_0)\,\gamma_5\,\gamma_k, \ (k=1,2,3)$ and using the ordinary nucleon 2-pt function $G_{\rm 2pt}(\vec{q},t_f,t_i,\Gamma_0)$ which is spin-projected by $\Gamma_0 = \frac{1}{4} \left( \idnty+\gamma_0 \right)\ $ we can define the ratio    
\vspace{-0.15cm}
\bea
{\rm R}^{\mu}_{\rm 3pt,{\cal Q}} \hspace{-0.05cm} \left(\hspace{-0.05cm}\vec{q}, t_f, t_i, t,\Gamma_k \right) \hspace{-0.1cm} = \hspace{-0.1cm} {\Large \frac{G^{\mu}_{\rm 3pt, \cal Q} \hspace{-0.05cm} (\vec{q}, t_f, t_i, t,
  \Gamma_k)}{G_{\rm 2pt} \hspace{-0.05cm} (\vec{q},t_f-t_i,\Gamma_0)} \hspace{-0.1cm} \sqrt{ \hspace{-0.1cm} \frac{G_{\rm 2pt}(\vec{q},t_f-t,\Gamma_0)G_{\rm
      2pt}(\vec{0},t-t_i,\Gamma_0)G_{\rm
      2pt}(\vec{0},t_f-t_i,\Gamma_0)}{G_{\rm 2pt}(\vec{0},t_f-t_i,\Gamma_0)G_{\rm 2pt}(\vec{q},t-t_i,\Gamma_0)G_{\rm 2pt}(\vec{q},t_f-t_i,\Gamma_0)}}}.
\label{eq:Ratio_3pt_2pt}
\vspace{-0.25cm}
\eea
As $t_f-t$, $t-t_i \to \infty$ we obtain the plateau $\Pi^{\mu}_{\rm 3pt,{\cal Q}} \left( \Gamma_k \right)= \lim_{t_f-t \to  \infty} \lim_{t-t_i \to \infty} {\rm R}^{\mu}_{\rm 3pt,{\cal Q}} \left(\vec{q}, t_f, t_i, t,\Gamma_k \right)$. By setting ${\vec{p}_f=0}$, $E_N=\sqrt{\vec{p}^2_i+m_N^2}$ and carrying out the Dirac algebra we obtain
\vspace{-0.15cm}
\bea
\Pi^{0}_{\rm 3pt,{\cal Q}} \left( \Gamma_k \right) = i\,C Q_k \left[ {\Large \frac{ \alpha^1 F_1(Q^2)}{2 m_N} + \frac{(E_N+3 m_N ) \alpha^1 F_2(Q^2)}{4 m^2_N} + \frac{ (E_N+m_N) F_3(Q^2)}{ 4m^2_N}} \right]\,, \label{eq:plateau_P0}
\vspace{-0.15cm}
\eea
where $C=({2m_N^2}/({E_N\left(E_N+m_N\right)}) )^{1/2}$ and  $\alpha^1$ denotes a phase due to $CP$-violation.  We  extract $\alpha^1$ using the plateau $\alpha^1 = \lim_{t_f-t_i \to \infty} {G_{\rm 2pt, {\cal Q}}
  \left(0, t_f, t_i,\Gamma_5 \right) } / {G_{\rm 2pt} \left(0, t_f, t_i, \Gamma_0 \right)}$ where ${G_{\rm 2pt, {\cal Q}}  \left(0, t_f, t_i,\Gamma_5 \right) }$ is a 2-pt function weighted by ${\cal Q}$ and spin-projected by $\Gamma_5={\gamma_5}/{4}$. Using~\eq{eq:plateau_P0} in combination with the plateaus leading to Sachs form factors \cite{Alexandrou:2014exa} $\Pi^{j}_\mathrm{3pt}(\Gamma_0) = \frac{-iC}{2\,m_N} Q_j \left( F_1(Q^2) - \frac{Q^2}{4m_N^2} F_2(Q^2) \right) \label{eq:q_G_E}$ and $\Pi^{j}_\mathrm{3pt}(\Gamma_k) = \frac{-C}{2\,m_N} \epsilon_{jik} Q_i \left( F_1(Q^2) + F_2(Q^2)  \right) \label{eq:q_G_M}$ we extract the following linear combination of ratios
\vspace{-0.25cm}
\begin{eqnarray}
 \Pi^k_{F_3} = \Pi^{0}_{\rm 3pt,\cal Q}(\Gamma_k) + 
i\,\alpha^1\,\Pi^{k}_\mathrm{3pt}\l(\Gamma_0\r) + 
\alpha^1\,\frac{1}{2}\sum_{i,j=1}^3\epsilon_{jki}\Pi^{j}_\mathrm{3pt}\l(\Gamma_i\r) =
{\large \frac{C (E_N + m_N)}{4\,m_N^2}}\,Q_k\, F_3(Q^2)\,,
\label{eq:plateaus}
\vspace{-0.25cm}
\end{eqnarray}
for which the decomposition only depends on the desired form factor $F_3(Q^2)$. 
\eq{eq:plateaus}, however, gives access to 
$Q_k F_3(Q^2)$ ($k{=}1,2,3$) hindering a direct evaluation of $F_3(0)$. Thus, we adopt
two approaches to extract $F_3(0)$: i) a dipole parametrization of $F_3(Q^2)$ 
in $Q^2$ and fit to extract the value at $Q^2=0$ and 
ii) position space methods~\cite{Alexandrou:2014exa,Alexandrou:2015spa}.
\section{Lattice Calculation}
\label{sec:lattice}
\vspace{-0.25cm}
 We extract the nEDM using the ensemble produced with $N_f=2+1+1$ twisted mass fermions and the Iwasaki gauge action at lattice spacing of $a\simeq0.082$~fm,  pion mass 373~MeV and a spatial lattice extent of $L/a=32$, and referred to as B55.32. In total we sampled over 4623 configurations, enabling us to reach high accuracy and to provide a reliable benchmark for the different methods.
\section{Topological Charge}
\label{sec:topological_charge}
We extract the field theoretic topological charge ${\cal Q}=\int d^4x\, q(x)\,,$ with an $O(a^4)$ improved (lattice) definition 
for $q(x)$ and damp the UV fluctuations using cooling and the gradient flow~\cite{Luscher:2010iy}. In addition we use the Wilson, Symanzik tree-level improved and Iwasaki actions for smoothing. Both techniques~\cite{Bonati:2014tqa,Alexandrou:2015yba} provide similar results on purely topological observables with equivalence realized by a rescaling between the gradient flow time $\tau$ and the number of cooling steps $n_c$~\cite{Bonati:2014tqa,Alexandrou:2015yba}. Based on Ref.~\cite{Luscher:2010iy} we read $F_3/ 2 m_N$ at a value of $\tau=t/a^2$ that satisfies the equality $\sqrt{8 t} \approx 0.6 {\rm fm}$. This corresponds~\cite{Alexandrou:2015spa} to $\tau$($n_c$) of 6.7(20) for Wilson, 7.1(30) for Symanzik tree-level improved and 6.3(50) for the Iwasaki action.
\section{Results for $F_3(0)$}
\label{results}
The first step towards the evaluation of $F_3(0)$ is the extraction of the phase $\alpha^1$. We use the various definitions for ${\cal Q}$ (cooling/gradient flow with different smoothing actions) to extract $\alpha^1$. Using ${\cal Q}$ measured via the gradient flow with the Iwasaki action yields $\alpha^1=-0.217(18)$ with plateau setting in at $t_f/a>8$ (we set $t_i=0$)~\cite{Alexandrou:2015spa}.  We subsequently use the values of $\alpha^1$ in \eq{eq:plateaus} in order to evaluate $F_3(0)$ with the different momentum treating techniques. \par
\subsection{Dipole Fit}
\label{sec:dipole_fit}
A technique to extract $F_3(0)$ from \eq{eq:plateaus} is the parametrization of the $Q^2$ dependence of $F_3 \left( Q^2 \right)$ and then fitting to determine $F_3(0)$ using the dipole form $F_3 \left( Q^2 \right) = {F_3(0)}/{\left( 1 + {Q^2}/{m_{F_3}^2} \right)^2}\,$. Hence, we compute $F_3(Q^2)$ for a sequence of values of the momentum transfer, $Q^2=2m_N \sqrt{E_N - m_N}$, with the spatial components $Q_i$ taking all possible combinations of $Q_i / (2\pi/L) \in [0,\pm 4]$ (and all permutations thereof). We perfrom the calculation at three source-sink separations $t_f=10a, 12a$ and $14a$. We report results for $t_f=12a$ since these are fully consistent with those for $t_f=14a$. In the left panel of \fig{fig:3pt_plateau} we show an example for the combination of ratios leading to the extraction of $F_3(Q^2)$. This corresponds to momentum transfer $Q^2\simeq0.17\, {\rm GeV}^2$ and ${\cal Q}$ extracted via gradient flow with the Iwasaki action. $F_3(Q^2)$ is extracted via a constant fit within the plateau region $t/a \in [3,9]$. In order to compute the error, we first calculate $F_3$ and the associated jackknife error $d F_3$ by employing the mean value for $\alpha^1$. We, then, recompute $F'_3$ using 
 $\alpha^1_{\rm max}=\alpha^1 + d\alpha^1$, where $d\alpha^1$ is the jackknife error of $\alpha^1$. The final error on $F_3$  is computed by combining $\Delta F_3 = F_3-F'_3$ due to the variation in $\alpha^1$  with the jackknife error $dF_3$ in quadrature, namely $\sqrt{(\Delta F_3)^2 + (d F_3)^2}$. After determining $F_3(Q^2)$ at each value of $Q^2$ we perform a dipole fit, treating $F_3(0)$ as a fitting parameter. For ${\cal Q}$ extacted via gradient flow and the Iwasaki action we report an nEDM of $F_3(0)/2 m_N = -0.041(12) \ {\rm e} \cdot {\rm fm}$. \par
\subsection{Position space methods}
\label{sec:position_space_methods}
The other momentum treating technique~\cite{Alexandrou:2014exa} is
based on removing the momentum factor in front of $F_3(Q^2)$ by position space methods. 
This involves the ``continuum derivative technique'' and ``momentum elimination technique''
which are briefly explained in the next two subsections. More information on these methods can be found in our longer write-up~\cite{Alexandrou:2015spa}.\par
\subsubsection{Continuum Derivative}
One can remove the
$Q_k$ dependence in front of $F_3(Q^2)$  by applying a continuum derivative with respect to $Q_j$ such as $\lim_{Q^2 \to 0} \frac{\partial}{\partial Q_j} \Pi^k_{F_3} ( \vec{Q} ) = \frac{C \left( E_N +m_N \right)}{4 m_N^2} \delta_{kj}  F_3(0)$. We explicitly show the application of the continuum derivative on the ratio in~\eq{eq:Ratio_3pt_2pt}, which leads to the first term in~\eq{eq:plateaus}; the generalization on the other two ratios is straightforward. This gives
\bea
\lim_{Q^2 \to 0} \frac{\partial}{\partial Q_j}  {\rm R}^{\mu}_{\rm
  3pt,{\cal Q}} \left(\vec{q}, t_f, t_i, t,\Gamma_k \right) = \lim_{Q^2 \to
  0}   \frac{\frac{\partial}{\partial Q_j}G^\mu_{{\rm 3pt}, {\cal Q}}
    (\vec{q}, t_f, t_i, t, \Gamma_k)}{G_{\rm 2pt}(\vec{q},t_f,t_i,\Gamma_{0})}\, =  \lim_{L \to \infty} \frac{\sum_{x = -L/2 + a}^{L/2 -
    a} i x_j G^\mu_{\rm 3pt, {\cal Q}} (\vec{x}, t_f, t_i, t, \Gamma_k)}{G_{\rm
      2pt}(\vec{q},t_f,t_i,\Gamma_{0})} \label{eq:fourier},
\eea
where the 3-pt function $ G^\mu_{{\rm 3pt},
  {\cal Q}} (\vec{x}, t_f, t_i, t, \Gamma_k)$ is expressed in position
space. In finite volume this expression
approximates the derivative of a $\delta$-distribution in momentum
space, 
\bea
a^3 \sum_{\vec{x}} i x_j G^\mu_{\rm 3pt, {\cal Q}} (\vec{x}, \Gamma_k) = \frac{1}{V} \hspace{-0.1cm} \sum_{\vec{k}} \hspace{-0.1cm} \left( \hspace{-0.1cm} a^3 \sum_{\vec{x}} \hspace{-0.1cm} i
x_j e^{\left( i \vec{k} \vec{x}  \right)}   \hspace{-0.1cm} \right) \hspace{-0.1cm} G^\mu_{{\rm 3pt}, {\cal Q}} (\vec{q}, \Gamma_k) \hspace{-0.1cm} \stackrel{L\rightarrow \infty}{\longrightarrow} \hspace{-0.1cm} \frac{1}{\left( 2 \pi
  \right)^3} \hspace{-0.1cm} \int \hspace{-0.1cm} d^3 \vec{k} \frac{\partial \delta^{(3)}(\vec{k})}{\partial k_j}
 G^\mu_{{\rm 3pt}, {\cal Q}} (\vec{k}, \Gamma_k). \label{eq:summation}   
\eea
For finite $L$ this implies a residual $t$-dependence $G^\mu_{{\rm 3pt}, {\cal Q}} (\vec{q}, t_f, t_i, t, \Gamma_k)
\sim {\rm exp} \left( -\Delta E_N t  \right)$ with $\Delta E_N = E_N \left( \vec{q} \right) - m_N$. Only  for $L \to \infty$ we have
$\Delta E_N \to 0$. 

Hence, the basic building blocks for this
technique are the standard 2-pt functions and the continuum
derivative-like 3-pt functions ${\partial}G^\mu_{{\rm 3pt}, {\cal Q}} (\vec{q}, t_f, t_i, t, \Gamma_k)/{\partial Q_j}$, 
${\partial}G^k_{{\rm 3pt}} (\vec{q}, t_f, t_i, t, \Gamma_0)/{\partial Q_j}$ as well as
${\partial}G^k_{{\rm 3pt}} (\vec{q}, t_f, t_i, t, \Gamma_i)/{\partial Q_j}$. These involve 3-pt functions in position space which are multiplied by $x_j$ in the final Fourier transform according to \eq{eq:fourier}. In addition, this method requires a sufficiently large cutoff for the summation in \eq{eq:summation} which needs to be checked explicitly. \par

\begin{figure}[t!]
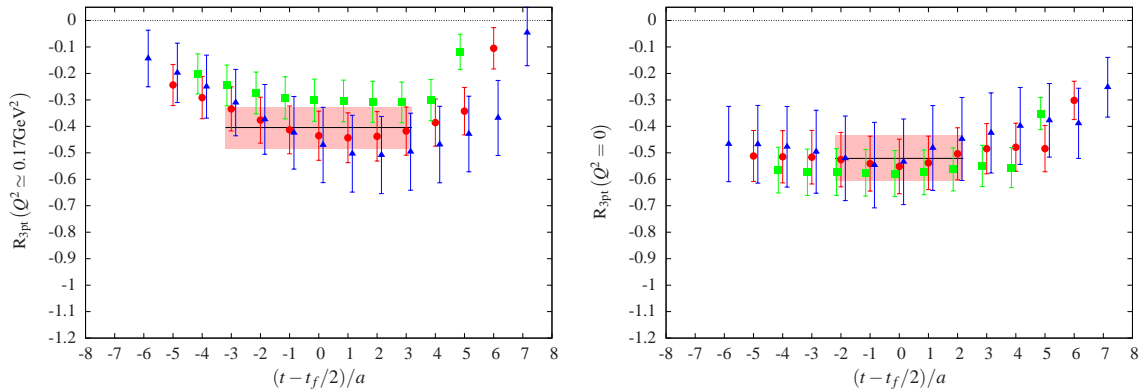

\vspace{-0.5cm}
\centerline{ \scalebox{0.60}{\input{F3_plateau_via_dipole_fit.tex}} \scalebox{0.60}{\input{F3_plateau_via_cont_der.tex}}}
\caption{\label{fig:3pt_plateau} The ratios leading to $F_3(Q^2\simeq0.17\, {\rm GeV}^2)$ (left panel) and $F_3(0)$ via continuum derivative (right panel). Green squares, red circles and blue triangles correspond
  to source-sink separation of $t_f/a=10,\,12,\,14$, respectively. ${\cal Q}$ is extracted using the gradient flow with the Iwasaki action. }
\vspace{-0.35cm}
\end{figure}

We, therefore, compute the right ratios and extract $F_3(0)/2 m_N$ at a source-sink separation of $t_f=12a$ and check for ground state dominance at $t_f=10a, 14a$. We fit the ratio to a constant in the plateau region to extract $F_3(0)$ and use the procedure explained in \sec{sec:dipole_fit} by employing the mean value for $\alpha^1$ as well as $\alpha^1_{\rm max}=\alpha^1 + d\alpha^1$ in order to compute the associated statistical error on $F_3$. Since, the final result has a residual time dependence of the form $\sim {\rm exp} \left(  a (E_N (\vec{q}) -m_N) \ t/a  \right)$ we perform an exponential fit to extract $F'_3$ and take the difference between $F_3$ and $F'_3$ as the systematic error. On the right panel of \fig{fig:3pt_plateau} we show results for the combination of continuum-like derivatives of ratios leading to the extraction of $F_3(0)$. The results are produced with ${\cal Q }$ extracted using the gradient flow with the Iwasaki action. We fit within the plateau range $t/a \in [4,8]$ yielding a value of $F_3(0)/2 m_N = -0.042(7)(3) \ {\rm e} \cdot {\rm fm}$.\par
\subsubsection{Momentum Elimination}
\label{sec:momentum_elimination}
In this method we start with a fit to the plateau in Eq.~(\ref{eq:plateaus}) and remove the time dependence. For simplicity we focus on the on-axis momenta, e.g. $\vec{q}=(\pm Q, 0,0)^T$. We average over all momentum directions and index combinations according to \eq{eq:plateaus} for a given $Q$-value. We denote the corresponding fitted ratios by $\Pi(Q)$. Subsequently, we apply a Fourier transform on $\Pi(Q)$ and obtain $\Pi(y)$ in position space by imposing a cutoff $Q_\mathrm{max}$. We exactly antisymmetrize~\cite{Alexandrou:2015spa} $\Pi(y)  \to \overline{\Pi}(n)$ with $n=y/a$ and Fourier transform it to continuous momentum space with $\Pi(k) =\l[ e^{ikn}\overline{\Pi}(n)\r]_{n=0,\,N/2} + 2i \sum_{n=1}^{N/2-1}  \overline{\Pi}(n) \sin\l(k\cdot n \r)$ and define the expressions $\hat{k} \equiv 2\sin\bigl(\frac{k}{2}\bigr)$ and $P_n\bigl(\hat{k}^2\bigr)=P_n\bigl(\bigl(2\sin\bigl(\frac{k}{2}\bigr)\bigr)^2\bigr) = \sin(nk) / \sin\bigl(\frac{k}{2}\bigr)$. This leads to $\Pi(\hat{k})-\Pi(0) = i \sum_{n=1}^{N/2-1} \hat{k}\, P_n \,\bigl(\hat{k}^2\bigr)  \overline{\Pi}(n)$. The function $P_n \,\bigl(\hat{k}^2\bigr)$ is related to Chebyshev polynomials of the second kind and it is analytic in $(-\infty, +4)$, allowing to evaluate $\Pi(\hat{k})$ at any intermediate value. Dropping the factor $\hat{k}$ in the above expression by differentiating, we obtain ${F_3(\hat{k}^2)}/{2m_N} = i\sum_{n=1}^{N/2-1} P_n(\hat{k}^2) \, \overline{\Pi}(n)$  without explicit momentum factors.
This expression can be computed exactly on the lattice up to the cutoff in the initial Fourier transform resulting in a smooth curve for $F_3(Q^2)$. 

We can extend the approach to arbitrary sets of
off-axis momentum classes  $M(Q,Q_\mathrm{off}^2) = \l\{\vec{q} \ | \ \vec{q}=\{\pm Q, Q_1, Q_2\} \,, \ Q_1^2+Q_2^2=Q_\mathrm{off}^2 \r\}\,$ where $\{\pm Q, Q_1, Q_2\}$ denotes all permutations of $\pm Q$, $Q_1$
and $Q_2$. To combine the results for $F_3(Q^2)$ for
different $Q_\mathrm{off}^2$--classes as a function of continuous momenta $Q^2=Q^2(\hat{k}, Q_\mathrm{off}^2)$ we need to consider an analytic continuation ($k\rightarrow i\kappa$ and $\hat{k} \rightarrow i\hat{\kappa} = -2\sinh\bigl({\kappa}/{2}\bigr)$) for classes with
$Q_\mathrm{off}^2>0$ to reach zero total momentum. This also affects $P_n$,
i.e. $P_n\bigl(\hat{\kappa}^2\bigr) = \sinh(n\kappa) /
\sinh\bigl(\frac{\kappa}{2}\bigr)$. The final result is obtained by combining the results from several sets of momentum classes $M(Q,Q_\mathrm{off}^2)$ by taking the error weighted average of the separate results. \par 
\begin{figure}[t!]
\vspace{-0.5cm}
\centerline{\hspace{0.0cm}\includegraphics[scale=0.35,angle=270]{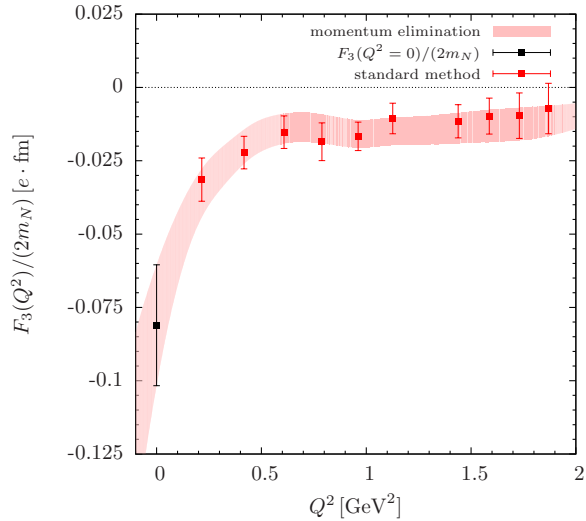}}
 \caption{\label{fig:ysummation} 
   Results on nEDM using the momentum
   elimination method for source-sink separation
   $t_{f}=12a$.}
\vspace{-0.25cm}
\end{figure}
Using this method we calculate $F_3(0)$ using various definitions for ${\cal Q}$. We analyze data for two source-sink separations, namely of $t_{f}=12a$ and $t_{f}=14a$, employing a general momentum cutoff $Q^2<16\cdot(2\pi/L)^2$ and momentum classes with an off-axis momentum squared of up to $Q^2_\mathrm{off} \leq 5 \cdot (2\pi/L)^2$. The red band in Fig.~\ref{fig:ysummation} shows the results for $F_3(Q^2)/(2m_N)$ extracted at $t_{f}=12a$ with ${\cal Q}$ obtained via the gradient flow with the Iwasaki action. It is obtained as the error
weighted average over all sets of different off-axis momentum classes $M(Q,Q_\mathrm{off}^2)$. As required, the band reproduces the red
points which are the results obtained using plateau method at each $Q^2$ value. Fig.~\ref{fig:ysummation} yields $F_3(0)/2 m_N = -0.082(21) \ {\rm e} \cdot {\rm fm}$. \par
\section{Conclusions}
\label{sec:conclusions}
\begin{figure}[t!]
\vspace{-1.5cm}
\centerline{\hspace{0.0cm}\includegraphics[scale=0.35,angle=270]{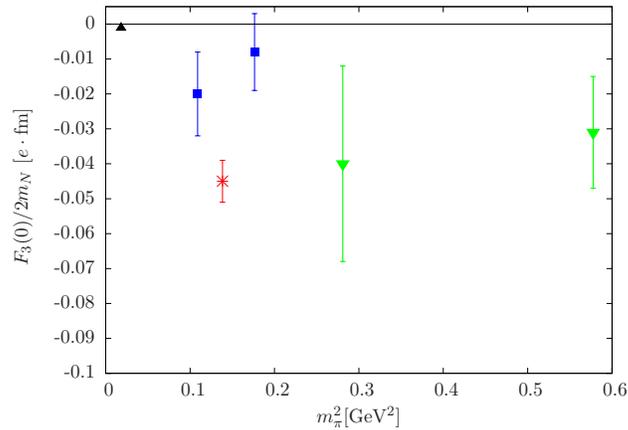}}
\vspace{-0.5cm}
\caption{\label{fig:Lattice_Comparison} $F_3(0)/(2 m_N)$
versus the pion mass squared ($m_\pi^2$). Our results are shown with a red asterisk. We also
show results for $N_f{=}2{+}1$
domain wall fermions~\cite{Shintani:2014zra} and $F_3(0)$ obtained with the usual parametrization in $Q^2$ (blue squares) as well as results for $N_f{=}2$ Clover fermions~\cite{Shintani:2008nt} extracted with the method of the background electric field (green triangles).  All errors shown are statistical. A value determined in chiral perturbation theory  is shown with the black  triangle~\cite{Ottnad:2009jw}.}
\end{figure}

The nEDM is computed using $N_f{=}2{+}1{+}1$ twisted mass fermions simulated at a pion mass of 373~MeV and lattice spacing of $a \simeq 0.082$~fm employing a total of 4623 measurements.
We observe~\cite{Alexandrou:2015spa} that different momentum treating techniques give similar results, with those extracted from the momentum elimination tending to have slightly larger
absolute values with larger errors. This resembles data obtained for the isovector magnetic form factor~\cite{Alexandrou:2014exa}. Results extracted using different actions to smooth the gauge links entering the computation of the topological charge yield overall consistent results~\cite{Alexandrou:2015spa}. Given that the simulations used the Iwasaki gauge action, we present as our final value for the nEDM the value extracted when the Iwasaki action is employed to obtain ${\cal Q}$.
As systematic error we take the difference between the mean values obtained when cooling and the gradient flow are used to determine ${\cal Q}$. Our final result is, thus, $F_3(0)/(2m_N) = -0.045(6)(1)\, e\cdot~{\rm fm}$ and it is in good agreement with older investigations~\cite{Alexandrou:2015spa} as this is demonstrated in~\fig{fig:Lattice_Comparison}. Using the experimental upper bound $\vert \vec{d}_N \vert = 2.9 \times 10^{-13} e \cdot {\rm fm}$ we find a maximum value of $\theta_{\rm max}=6.4(0.9)(0.2)\times 10^{-12}$. Further information on this work can be found in our longer write-up~\cite{Alexandrou:2015spa}. 
\section*{Acknowledgments}
Numerical calculations were carried out in HPC resources from John von  Neumann-Institute for Computing on the JUQUEEN and JUROPA systems at the research  center in J\"ulich as well as by the Cy-Tera machine at The Cyprus Institute. This work was supported, in part, by a grant from the Swiss National Supercomputing Centre (CSCS) under project ID s540.

\end{document}